\documentclass[a4paper,12pt]{article}
\usepackage[english]{babel}
\usepackage{amsfonts,amssymb,amsmath}
\begin{document}

 \centerline{ \textbf{ \Large ASYMPTOTICS OF SOLUTIONS }}
 \centerline{ \textbf{ \Large  OF A PARABOLIC EQUATION } }
 \centerline{ \textbf{ \Large NEAR SINGULAR POINTS} }

 \vskip 5mm 
 \centerline{SERGEI V.~ZAKHAROV}

 \vskip 10mm 
\noindent
ABSTRACT.
Results of investigation of the asymptotic behavior
of solutions to the Cauchy problems
for a quasi-linear parabolic equation with a small parameter at
a higher derivative near singular points
 of limit  solutions are presented. 
Interest to the problem under consideration is explained
by its applications to a wide class of physical systems and
probabilistic processes such as acoustic waves in fluid and gas,
hydrodynamical turbulence and nonlinear diffusion. 
The following cases  are considered: 
a singularity generated by a jump discontinuity 
of the initial function, collision of two shock waves,
gradient catastrophe, transition of a weak discontinuity
into a  shock wave, a singularity generated by a large 
initial gradient.

 \section{Introduction.}

A simplest model of the motion of continuum,
which takes into account nonlinear effects and dissipation,
 is the equation of nonlinear diffusion $$ 
 \frac{\partial u}{\partial t} + 
 u\frac{\partial u}{\partial x} = 
 \varepsilon \frac{\partial^2 u}{\partial
x^2} 
 $$ 
for the first time presented by J.~Burgers~\cite{bu}. 
 This equation  is used in studying
 the evolution of a wide class of physical systems and probabilistic
process, acoustic waves in fluid and gas~\cite{wh}.

 In the present survey, results of investigations
of the asymptotic  behavior of solutions to
the Cauchy problems for a more general quasi-linear parabolic
equation  
 \begin{eqnarray} 
 \label{eq} 
 \frac{\partial u}{\partial t} + 
 \frac{\partial \varphi(u)}{\partial
x} = 
 \varepsilon \frac{\partial^2 u}{\partial
x^2}, 
 & \quad t\geqslant t_0, \\ 
 \label{ic}\phantom{\frac{1}{1}} 
 u(x,t_0) = q(x), & \quad x\in\mathbb{R},
 \end{eqnarray} 
are collected.
 We assume that $\varepsilon>0,$ 
the function $\varphi$ is infinitely differentiable
 and its second derivative is strictly positive. 
 The initial function~$q$ is bounded and piecewise smooth. 
 
 Such models are constructed for studying 
the motion of motor transport, flood waves 
and others~\cite{ib}. 
The interest to problem under consideration is explained
by physical applications and 
 the fact that its solutions allow one to obtain viscous 
generalized solutions of  the limit equation. 
This problem had been studied by
N.S.~Bakhvalov, I.M.~Gelfand, A.M.~Il'in, 
E.~Hopf, O.A.~Ladyzhenskaya, O.A.~Oleinik,
 and many other mathematicians. 
It is strictly proved~\cite{lsu}, 
 that there exists a unique bounded infinitely
differentiable with respect to $x$ and~$t$
solution $u(x,t,\varepsilon)$.

 In this survey, we present results
of investigation of the asymptotic behavior of solutions
to the Cauchy problems in neighborhoods
of singular points, which arise on discontinuity sets
of the limit (degenerate) solution with $\varepsilon=0$. 
 In the next sections the following problems
 are considered.

 1. {\it Jump discontinuity of the initial function.} 
 A.M.~Il'in and T.N.~Nesterova~\cite{in2} 
 considered problem~(\ref{eq})--(\ref{ic}) in the case when the initial
function has a finite jump discontinuity. 
  
 2. {\it Collision of two shock waves.} 
 In the same paper~\cite{in2}, problem~(\ref{eq})--(\ref{ic}) is considered in the
case when the limit solution on a finite interval
of time has two smooth curves of discontinuity
 \mbox{$x=s_1(t)$} and \mbox{$x=s_2(t)$,} 
 merging at the moment $t=t^{*}$ into
one \mbox{$x=s_3(t)$}. 
 For the Burgers equation this case is given 
in  Whitham's book~\cite{wh}. 
 
3. {\it Gradient catastrophe.} 
 In~\cite{ilyn}, A.M.~Il'in  studied the
case when in the strip $\{(x,t):$$~t_0\leqslant t\leqslant T,$$~x\in\mathbb{R}\}$ 
the limit ($\varepsilon=0$) solution of the problem is
 smooth everywhere except for one smooth discontinuity curve. 
 This problem in detail is considered
 in his classical monograph~\cite{ib}. 
 
 4. {\it Transition of a weak discontinuity into a shock wave.} 
 In~\cite{sushko}, V.G.~Sushko   studied
 problem~(\ref{eq})--(\ref{ic})  in the case
when the initial function $q(x)$ 
 is smooth everywhere except for one point, at which
it is continuous, while the first derivative
has a jump discontinuity. 
 Such a weak discontinuity in the limit problem propagates
along a characteristic line for a finite time
and then becomes a shock wave. 
The asymptotics of the solution
 in a small parameter $\varepsilon$ 
 in a neighborhood of the transition point
is obtained in~\cite{iz,gc}. 
 
 5. {\it Singularity generated by a large initial gradient.} 
 In papers~\cite{2ps,zz}, the author studied
the  problem with two small parameters: 
a viscosity (diffusion)  parameter $\varepsilon$ 
 at a higher derivative and an additional
parameter $\rho$ in the initial condition $u(x,t_0) = \nu (x/\rho)$. 
 
 Application of the matching method 
to investigation of solutions of the above problems
leads to the necessity of constructing
several asymptotic series in distinct subdomains
of independent variables. 
For this reason, these asymptotics are called singular
 and such problems  are classified as singularly perturbed.

 \setcounter{equation}{0} 
 \section{Discontinuity of the initial function.} 
 In papers by A.M. Il'in and T.N. Nesterova~\cite{in2},
the problem is considered in the case when the initial
function has a finite jump discontinuity. 
Assuming, without loss of generality, that the jump
of the function $u_0$ lies at the point $x=0$
and the initial moment of time is $t_0=0$, 
it is convenient to introduce the inner variables 
$$ 
 \zeta=\frac{x-s(t)}{\varepsilon}, 
 \quad 
 \tau =\frac{t}{\varepsilon}, 
 $$ 
 where $x=s(t)$ is a curve of discontinuity of the limit solution. 
Then equation~(\ref{eq}) becomes $$ 
 \frac{\partial^2 w}{\partial \zeta^2} 
 + s'(\varepsilon\tau)\frac{\partial
w}{\partial \zeta} 
 - \frac{\partial \varphi(w)}{\partial \zeta} 
 -\frac{\partial w}{\partial \tau}=0. 
 $$ 
 The asymptotics of the solution in a neighborhood
 of the singular point ($x=0$, $t=0$) is constructed in the form
of the series $$ 
 w = \sum\limits_{n=0}^{\infty} \varepsilon^{n}
w_n(\zeta,\tau), 
 $$ 
 whose coefficients are found from the system
of equations $$ 
 \frac{\partial^2 w_0}{\partial \zeta^2} 
 + s'(0)\frac{\partial w_0}{\partial \zeta} 
 - \frac{\partial \varphi(w_0)}{\partial \zeta} 
 -\frac{\partial w_0}{\partial \tau}=0, 
 $$ 
 $$ 
 \frac{\partial^2 w_n}{\partial \zeta^2} 
 + \frac{\partial [(s'(0)-\varphi'(w_0))w_n]}{\partial \zeta} 
 -\frac{\partial w_n}{\partial \tau}= 
 $$ 
 $$ 
 =\frac{\partial}{\partial \zeta} 
 G_n(w_0,\dots,w_{n-1}) 
 -\sum\limits_{j=1}^{n}\frac{\tau^j}{j!} 
 \frac{d^{j+1}s(0)}{dt^{j+1}} 
 \frac{\partial w_{n-j}}{\partial \zeta}, 
 \quad n\geqslant 1,
 $$ 
with the following the initial conditions: 
$$
 w_n(\zeta,0) = 
 \begin{cases} 
 \displaystyle\frac{1}{n!}\frac{d^{n}u_0(+0)}{dx^{n}}\zeta^n, & \zeta>0, \\ 
 \displaystyle\frac{1}{n!}\frac{d^{n}u_0(-0)}{dx^{n}}\zeta^n, & \zeta< 0. 
 \end{cases} 
$$
 It is proved that there exist infinitely
differentiable for $|\zeta|+\tau>0$ solutions of this
system of initial value problems.

 \setcounter{equation}{0} 
 
 \section{Collision of two shock waves.}
Also in~\cite{in2}, 
the problem is considered in the case when
the limit solution on a finite interval  of time has two smooth
curves of discontinuity \mbox{$x=s_1(t)$} and \mbox{$x=s_2(t)$,} 
 merging at the moment $t=t^{*}$ into one curve \mbox{$x=s_3(t)$}. 
 For the solution the junction point is also singular;
 and in its neighborhood it is necessary to introduce
the inner variables $$ 
 \zeta=\frac{x-x^{*}}{\varepsilon}, 
 \quad 
 \tau =\frac{t-t^{*}}{\varepsilon},
 $$ 
where $x^{*}=s_3(t^{*})$.
Then equation~(\ref{eq}) becomes $$ 
 \frac{\partial w}{\partial \tau} 
 -\frac{\partial^2 w}{\partial \zeta^2} 
 + \frac{\partial \varphi(w)}{\partial \zeta} =0. 
 $$ 
 The asymptotics of the solution in a neighborhood 
of the singular point $(x^{*},t^{*})$ is constructed in the form
of the series $$ 
 w = \sum\limits_{n=0}^{\infty} \varepsilon^{n}
w_n(\zeta,\tau), 
 $$ 
 whose coefficients are found from the system of 
equations $$ 
 \frac{\partial^2 w_0}{\partial \zeta^2} 
 - \frac{\partial \varphi(w_0)}{\partial \zeta} 
 -\frac{\partial w_0}{\partial \tau}=0, 
 $$ 
 $$ 
 \frac{\partial^2 w_n}{\partial \zeta^2} 
 - \frac{\partial [\varphi'(w_0)w_n]}{\partial \zeta} 
 -\frac{\partial w_n}{\partial \tau}= 
 \frac{\partial}{\partial \zeta} 
 g_n(w_0,\dots,w_{n-1}), 
 \quad n\geqslant 1. 
 $$ 
 These equations should be supplied with conditions
 which are obtained as follows.
The composite asymptotics, approximating the solution
for $t<t^{*}$, is rewritten in the inner variables:
 $$ 
 U_N(x,t,\varepsilon) = \sum\limits_{n=0}^{N} 
 \varepsilon^{n} S_n(\zeta,\tau)+ 
 $$ 
 $$ 
 + O\left\{ \varepsilon^{n+1} \left[ 1 + 
 | \tau|^{2(n+1)}\left( |\zeta - s'_1(t^{*})\tau|^{n} 
 + |\zeta - s'_2(t^{*})\tau|^{n} 
 \right) \right] 
 \right\}, 
 $$ 
 where functions $S_n(\zeta,\tau)$ are polynomials of
degree $2n$ in $\tau$. 
 Thus, it is required the fulfillment of the
conditions $$ 
 \lim\limits_{\tau\to -\infty} 
 (w_n(\zeta,\tau)-S_n(\zeta,\tau))=0. 
 $$ 
 It is proved that solutions $w_n$ exist and
satisfy the estimates $$ 
 \Bigg|\frac{\partial^{l+m} }{\partial \zeta^l\partial\tau^m} 
 (w_n(\zeta,\tau)-S_n(\zeta,\tau)) 
 \Bigg| < M \exp \{ \mu\tau- \gamma|\zeta|\}, 
 \qquad 
 \tau<0, 
 $$ 
 $$ 
 \Bigg|\frac{\partial^{l+m} }{\partial \zeta^l\partial\tau^m} 
 (w_n(\zeta,\tau)-R_{3,n}(\zeta-s'_3(t^{*})\tau,\tau)) 
 \Bigg| < M \exp \{ - \gamma(\tau+|\zeta|)\}, 
 \qquad 
 \tau>0, 
 $$ 
 where~$M$, $\mu$, $\gamma$ are positive constants
depending on $k$, $l$ and~$m$; 
 $R_{3,n}(\zeta,\tau)$ are polynomials in $\tau$ 
 obtained from reexpansion of the asymptotics $$ 
 \sum\limits_{k=0}^{\infty} 
 \varepsilon^{k} v_{3,k}((x-s_3(t))/\varepsilon,t)= 
 \sum\limits_{k=0}^{\infty} 
 \varepsilon^{k} R_{3,k}(\zeta-s'_3(t^{*})\tau,\tau)) 
 $$ 
 in a neighborhood of the shock wave $x=s_3(t)$.

 \vskip 10mm 
 \setcounter{equation}{0} 
 
 \section{Gradient  catastrophe.} 
 A.M. Il'in studied the case~\cite{ilyn}
 when in the strip $$\{(x,t):~t_0\leqslant t\leqslant T,~x\in\mathbb{R}\}$$ 
the limit ($\varepsilon=0$) solution of the problem is
a smooth  function everywhere except for one
smooth discontinuity curve $$ 
 \{ (x,t) : x= s(t),\ t\geqslant t^{*}>
t_0\}. 
 $$ 
A detailed presentation of his results can be found in
monograph~\cite{ib}, 
 where the asymptotics of the solution
as $\varepsilon\to 0$ is constructed and justified with
an arbitrary accuracy. 
Under a suitable choice of independent variables,
the  singular point $(s(t^{*}),t^{*})$ 
coincides with the origin and
 in its neighborhood 
the following stretched variables are introduced:
 $$ 
 \xi = \varepsilon^{-3/4} x, 
 \quad 
 \tau = \varepsilon^{-1/2} t.
 $$ 
An expansion of the solution is sought in the form
of the series
 $$ 
 w = \sum\limits_{k=1}^{\infty} \varepsilon^{k/4} 
 \sum\limits_{j=0}^{k-1} w_{k,j}(\xi,\tau) 
 \ln^{j} \varepsilon^{1/4}.
 $$ 
Substituting it into equation~(\ref{eq}), 
for coefficients $w_{k,j}$ we obtain
the recurrence system $$ 
 \frac{\partial w_{1,0}}{\partial \tau} 
 + \varphi''(0) w_{1,0}\frac{\partial w_{1,0}}{\partial \xi} 
 -\frac{\partial^2 w_{1,0}}{\partial \xi^2} =0, 
 $$ 
 $$ 
 \frac{\partial w_{k,j}}{\partial \tau} 
 + \varphi''(0) \frac{(\partial w_{1,0} w_{k,j})}{\partial \xi} 
 -\frac{\partial^2 w_{k,j}}{\partial \xi^2} =E_{k,j}, 
 \quad k\geqslant 2.
 $$ 
 These equations should be supplied with the conditions $$ 
 w_{k,j}(\xi,\tau) =W_{k,j}(\xi,\tau), 
 \quad \tau\to -\infty, 
 $$ 
 where $W_{k,j}(\xi,\tau)$ is the sum of all
coefficients at $\varepsilon^{k/4}\ln^{j} \varepsilon^{1/4}$
 in the reexpansion of the asymptotics far from
the singularity (the outer expansion) 
 in terms of the inner variables. 
 
 Investigation of solutions of this system is
a central and the most laborious task in this problem. 
 It is proved that there exist solutions $w_{k,j}(\xi,\tau)$ 
 for $k\geqslant 2$, $0 \leqslant j \leqslant
k-1$, 
 infinitely differentiable for all $\xi$ and $\tau$.

 Observe separately the properties of the leading term,
 which is found with the help of the Cole--Hopf transform $$ 
 w_{1,0} (\xi,\tau) = - \frac{2}{\varphi''(0)\Lambda(\xi,\tau)}
 \frac{\partial \Lambda(\xi,\tau)}{\partial \xi}, 
 $$ 
 where $$ 
 \Lambda(\xi,\tau) =\int\limits_{-\infty}^{\infty} 
 \exp\left( -\frac{1}{8} 
 (z^4 - 2 z^2 \tau +4 z \xi) 
 \right) dz 
 $$ 
 is a solution of the heat equation. 
The function $w_{1,0}$ satisfies the condition $$ 
 w_{1,0} (\xi,\tau) = [\varphi''(0)]^{-1} H(\xi,\tau) + 
 \sum\limits_{l=1}^{\infty} h_{1-4l}(\xi,\tau), 
 \qquad 
 3[H(\xi,\tau)]^2-\tau\to \infty, 
 $$ 
 where $H(\xi,\tau)$ is the Whitney fold
function, 
$$H^3 - \tau H + \xi = 0,$$ 
 $h_l(\xi,\tau)$ are homogeneous functions
of power~$l$ 
in  $H(\xi,\tau)$, $\sqrt{-\tau}$
and $\sqrt{3[H(\xi,\tau)]^2-\tau}$, 
 which are polynomials in $H(\xi,\tau)$, $\tau$
and $(3[H(\xi,\tau)]^2-\tau)^{-1}$. 
 
 In addition, there hold the following formulas: 
 $$ 
 w_{1,0} (\xi,\tau) = |\tau|^{1/2} 
 \left(Z_0(\theta)+ 
 \sum\limits_{j=1}^{\infty} 
 \tau^{-2j }Z_j(\theta) 
 \right), 
 \qquad 
 \tau \to-\infty, 
 $$ 
 where $\theta= \xi|\tau|^{-3/2}$, 
 $Z_j\in C^{\infty}(\mathbb{R}^1)$ are solutions of
a recurrence system of ordinary differential
equations; 
 $$ 
 w_{1,0}(\xi,\tau) = \sqrt{\tau} 
 \left( -\frac{\mathrm{th}\, z}{\varphi''(0)} 
 +\sum\limits_{k=1}^{\infty} \tau^{-2k}
q_k(z) 
 \right), 
 \qquad 
 \tau \to+\infty, 
 \quad 
 |\xi| \tau^{1/2}< \tau^{\alpha}, 
 $$ 
 where $z= \xi\sqrt{\tau}/2$, $\alpha > 0$,
 $q_k\in C^{\infty}(\mathbb{R}^1)$ are solutions 
of a recurrence system of ordinary differential
equations. 
 
 \vskip 10mm 
 \setcounter{equation}{0} 
 
 \section{Transition of a weak discontinuity into a shock wave.} 
 In the paper by V.G.Sushko~\cite{sushko},
 problem~(\ref{eq})--(\ref{ic}) is studied in the case
when the initial function $u(x,0,\varepsilon)$ 
 is smooth everywhere except for one point
 at which it is continuous and 
the first derivative has a jump discontinuity. 
 Then in some strip $t_0\leqslant t \leqslant
t^{*}$ 
the limit solution $u(x,t,0)$ is continuous;
 however, the derivative  $u_x$ has a jump
discontinuity, i.e., a weak  discontinuity. 
 In papers~\cite{iz,gc},
  the behavior of the solution $u(x,t,\varepsilon)$ 
 with the initial function $$ 
 -(x + ax^2)\,\Theta(-x)\,(1 + q_0(x)) 
 $$ 
is studied near the  transition  point,
 $\Theta(x)$ is the Heaviside function. 

 We assume that $\varphi\in C^\infty(\mathbb{R})$, $\varphi''(u)>0$, 
 $\varphi(0) = \varphi'(0) = 0$, $\varphi''(0) = 1$, 
 $\varepsilon>0$, $a>0$, 
 $q_0\in C^\infty(\mathbb{R})$, 
 $q_0(x)=0$ in some neighborhood of zero.

 In a neighborhood of the origin ($x=0,~t=0$),
 let us introduce the stretched variables $$ 
 \xi =\varepsilon^{-2/3}x, 
 \qquad 
 \tau =\varepsilon^{-1/3}t. 
 $$ 
 The asymptotics of the solution of the problem
 in a neighborhood of the origin is constructed
in the form of the series $$ 
 W = \sum\limits_{p=2}^{\infty}\varepsilon^{p/6} 
 \sum\limits_{s=0}^{[p/2]-1}\ln^s\varepsilon\, 
 w_{p,s}(\xi,\tau). 
 $$ 
 Coefficients $w_{p,s}(\xi,\tau)$ are solutions of
the recurrence system $$ 
 	\frac{\partial w_{2,0}}{\partial \tau} 
 	+w_{2,0}\frac{\partial w_{2,0}}{\partial \xi} 
 	-\frac{\partial^2w_{2,0}}{\partial \xi^2}=0, 
 $$ 
 $$ 
 	\frac{\partial w_{3,0}}{\partial \tau} + 
 	\frac{\partial \left(w_{2,0} w_{3,0}\right) }{\partial \xi} 
 	- \frac{\partial^2 w_{3,0}}{\partial \xi^2} = 0, 
 $$ 
 $$ 
 	\frac{\partial w_{p,s}}{\partial \tau} + 
 	\frac{\partial \left(w_{2,0} w_{p,s}\right) }{\partial \xi} 
 	- \frac{\partial^2 w_{p,s}}{\partial \xi^2} = 
 \frac{\partial E_{p,s}}{\partial \xi}, 
 $$ 
 where $$ 
 	E_{p,s} = - \frac{1}{2} 
 \sum\limits_{m=3}^{p-1} 
 \sum\limits_{l=0}^{s} 
 w_{m,l} w_{p+2-m,s-l} 
 -\sum\limits_{q=3}^{[p/2]-s+1}\frac{\varphi^{(q)}(0)}{q!} 
 	\sum\limits_{p_1+ \dots + p_q = p + 2 \atop
s_1+\dots +s_q = s} 
 \prod\limits_{j=1}^{q} w_{p_j,s_j} 
 $$ 
 (for $ s = [p/2] - 1 $ the sum in $q$ is equal to zero), 
 with conditions as $\tau\to -\infty$ 
 $$ 
 w_{p,s} = \sum\limits_{l = s}^{[p/2]-1} 
 \frac{l!}{s!(l-s)!3^s} 
 \ln^{l - s} |\tau| \sum\limits_{k = \max\{1,2l\}}^{\infty} 
 |\tau|^{(p - 3k)/2}R_{k,l,p - 2l-2}(\theta)
 $$ 
 in the domain $$ 
 X^0 = \{ (\xi,\tau):~|\xi| < |\tau|^{1 - \gamma},\ \tau < 0 \} 
 \qquad 
 (0 < \gamma < 1/2),
 $$ 
 functions $R_{k,l,p - 2l-2}(\theta)$ 
 are found from the matching condition of the 
series~$W$ 
 with the expansion in the boundary layer
near the line of a weak discontinuity. 
 Here, we use the notation for the self-similar
variable $$ 
 \theta = \frac{\xi}{2\sqrt{-\tau}}. 
 $$ 
 The leading term of the asymptotics has the
form $\varepsilon^{1/3}w_{2,0}(\xi,\tau)$. 
The function $w_{2,0}$ is defined by formulas
$$ 
 w_{2,0}(\xi,\tau)= 
 -\frac{2}{\Phi(\xi,\tau)} 
 \frac{\partial \Phi(\xi,\tau)}{\partial \xi}, 
 $$ 
 $$ 
 \Phi(\xi,\tau) = \int\limits_{0}^{\infty} 
 \exp\left( -\frac{4b}{3}s^3 + \tau s^2 - \xi s\right)
ds, 
 \qquad 
 b=a-\varphi'''(0)/2>0. 
 $$ 
The next coefficient of the expansion
is also obtained  in an explicit form:
 $$ 
 w_{3,0}(\xi,\tau) = 
 \frac{\sqrt{\pi}}{[\Phi(\xi,\tau)]^{2}} 
 \frac{\partial \Phi(\xi,\tau)}{\partial \xi}. 
 $$

The arrangement of boundary layers breaks
an analytic character of the asymptotics $W$
 at infinity that does not allow one to guess
solutions of the ``scattering problem'' for all
coefficients of the asymptotics, 
as was made in the problem with a smooth initial function. 

 \vskip 10mm 
 
 \section{Singularity generated by a large initial gradient.} 
 
 Another type of a singular point of  the solution
 arises in the case  with two small parameters~\cite{2ps}, 
 when the initial condition has the
form $$ 
 \phantom{\frac{1}{1}} 
 u(x,0,\varepsilon,\rho) = \nu ( {x}{\rho}^{-1}), 
 \quad x\in\mathbb{R}, \quad \rho>0, 
 $$ 
 where function $\nu$ is infinitely differentiable and bounded,
 and $\rho$ is the second small parameter. 
 In~\cite{zz}, it is proved that in this case
under conditions $$ 
 \nu(\sigma) = \sum\limits_{n=0}^{\infty} 
 \frac{\nu^{\pm}_n}{\sigma^{n}}, 
 \qquad 
 \sigma\to \pm\infty, 
 \qquad 
 (\nu^-_0>\nu^+_0) 
 $$ 
 for the solution of problem $(\ref{eq})$--$(\ref{ic})$ 
 as $\varepsilon\to 0$ and $\mu=\rho/\varepsilon\to 0$ 
 in the strip $$ 
 \{ (x,t) : x\in\mathbb{R},\ 0 \leqslant
t\leqslant T\} 
 $$ 
 there holds the asymptotic formula $$ 
 {u}(x,t,\varepsilon,\rho)= 
 h_0\left(  \frac{x}{\rho}, 
 \frac{\varepsilon t}{\rho^2}\right)
- R_{0,0,0}\left( 
 \frac{x}{2\sqrt{\varepsilon t}} 
 \right)+ 
 \Gamma\left( 
 \frac{x}{\varepsilon}, 
 \frac{t}{\varepsilon} 
 \right) 
 +O\left(\mu^{1/2}\ln\mu\right), 
 $$ 
 where $$ 
 h_0(\sigma,\omega) = 
 \frac{1}{2\sqrt{\pi\omega}} 
 \int\limits_{-\infty}^{\infty} 
 \nu(s) \exp\left[ -\frac{(\sigma-s)^2}{4\omega}\right] 
 ds, 
 $$ 
 $$ 
 R_{0,0,0}(z) = \nu^{-}_{0} 
 \mathrm{erfc}(z) + \nu^{+}_{0} \mathrm{erfc}(-z), 
 $$ 
 $$ 
 \mathrm{erfc} (z) = \frac{1}{\sqrt{\pi}} 
 \int\limits_{z}^{+\infty} \exp(-y^2)\,
dy, 
$$
$$
\sigma = \frac{x}{\rho},
 \qquad 
\omega = \frac{\varepsilon t}{\rho^2},
 \qquad 
 z = \frac{\sigma}{2\sqrt{\omega}}, 
 $$ 
the function $\Gamma$ is the solution 
of the equation in the inner variables ($\eta =
x/\varepsilon$, $\theta = t/\varepsilon$) 
 $$ 
 \frac{\partial \Gamma}{\partial \theta} + 
 \frac{\partial \varphi(\Gamma)}{\partial \eta} - \frac{\partial^2 \Gamma}{\partial \eta^2} = 0 
 $$ 
 with the initial condition $$ 
 \Gamma(\eta,0) = 
 \begin{cases} 
 \nu^-_0, & \eta<0, \\ 
 \nu^+_0, & \eta > 0. 
 \end{cases} 
 $$ 

 In addition, using the renormalization  method
the following asymptotic  formula 
is obtained: $$ 
 {u}(x,t,\varepsilon,\rho)= 
 \frac{1}{\nu^+_0 - \nu^-_0} 
 \int\limits_{-\infty}^{\infty} 
 \Gamma\left( \frac{x-\rho s}{\varepsilon}, \frac{t}{\varepsilon}\right) 
 \nu'(s)\, ds 
 +O\left(\mu^{1/4}\right). 
 $$ 
 These results give asymptotics only in the leading approximation. 

To construct a complete expansion near 
the singular point $x=0$, $t=0$,
it is natural to ``stretch'' the variable~$x$ on the value $\rho^{-1}$. 
To keep the evolutionary character of equation~(\ref{eq}),
the derivative with respect to $t$ must
be the same order as the right-hand side,
 i.e., of order $\varepsilon\rho^{-2}$. 
Thus, we make the change of 
variables $$ 
 x = \rho\sigma, 
 \qquad 
 t = \frac{\rho^2}{\varepsilon} \omega. 
 $$ 
 The inner expansion is sought in the form of the 
series $$ 
 H(\sigma,\omega,\mu)=\sum\limits_{n=0}^{\infty} 
 \mu^{n} h_n(\sigma,\omega), 
 \qquad 
 \mu = \frac{\rho}{\varepsilon} 
 \to 0.
 $$ 
Substituting it into the equation $$ 
 \frac{\partial h}{\partial \omega} - \frac{\partial^2
h}{\partial \sigma^2} = 
 - \mu \frac{\partial \varphi(h)}{\partial \sigma}, 
 $$ 
 for $h(\sigma,\omega)\equiv u(\rho\sigma,\rho^2 \omega/\varepsilon)$, 
 we obtain a recurrence chain of the initial value
problems $$ 
 \frac{\partial h_0}{\partial \omega}-\frac{\partial^2
h_0}{\partial \sigma^2} = 0, 
 \qquad h_0(\sigma,0) =\nu(\sigma), 
 $$ 
 $$ 
 \frac{\partial h_1}{\partial \omega}-\frac{\partial^2
h_1}{\partial \sigma^2} = - \frac{\partial \varphi(h_0)}{\partial \sigma}, 
 \qquad h_1(\sigma,0) = 0, 
 $$ 
 $$ 
 \frac{\partial h_{n}}{\partial \omega}-\frac{\partial^2 h_{n}}{\partial \sigma^2} 
 = - \frac{\partial E_{n}}{\partial \sigma}, 
 \qquad h_{n}(\sigma,0) = 0, 
 $$ 
 where $$ 
 E_{n} = 
 \sum\limits_{q=1}^{n-1} 
 \frac{\varphi^{(q)}(h_0)}{q!} 
 \sum\limits_{n_1+\dotsc+n_q=n-1} 
 \prod\limits_{p=1}^{q} h_{n_p}, 
 \qquad 
 n\geqslant 2. 
 $$ 
 It follows that all coefficients $h_n(\sigma,\omega)$ 
 are uniquely determined: 
 $$ 
 h_0(\sigma,\omega) = 
 \frac{1}{2\sqrt{\pi\omega}} 
 \int\limits_{-\infty}^{\infty} 
 \nu(s) \exp\left[ -\frac{(\sigma-s)^2}{4\omega}\right] 
 ds, 
 $$ 
 $$ 
 h_n(\sigma,\omega) = - \int\limits_{0}^{\omega} 
 \int\limits_{-\infty}^{\infty} 
 \frac{1}{2\sqrt{\pi(\omega-v)}} 
 \exp\left[ -\frac{(\sigma-s)^2}{4(\omega-v)}\right] 
 \frac{\partial E_n}{\partial s} 
 ds dv. 
 $$

\bigskip
\centerline{ACKNOWLEDGMENTS}
\smallskip

This work was supported by the Russian Foundation for Basic Research,
project no.~14-01-00322.

 \vskip 5mm 
\noindent
INSTITUTE OF MATHEMATICS AND MECHANICS,\\
URAL BRANCH OF THE RUSSIAN ACADEMY OF SCIENCES,\\
16, S.KOVALEVSKAJA STREET, 620219, EKATERINBURG GSP-384, RUSSIA 

 \vskip 5mm

\textit{E-mail address}: svz@imm.uran.ru

\end{document}